\documentclass[%
 reprint, 
 amsmath,amssymb,
 aps,
prx,
]{revtex4-2}
\usepackage{graphicx}
\usepackage{dcolumn}
\usepackage{bm}
\usepackage{amsmath, amssymb, braket, multirow, booktabs, array}
\usepackage{xcolor, tikz}
\DeclareMathOperator{\Tr}{Tr}

\usetikzlibrary{calc}
\begin{document}


\title{Measurements generate nonclassicality in noisy quantum first-passage dynamics}%
\title{Emergent quantum effects in monitored oscillator timing statistics}
\title{Quantum signatures in noisy monitored oscillators}
\title{Quantum-to-classical crossover in first-passage statistics of monitored noisy oscillators}
\title{First-passage dynamics and nonclassical conditioned states in a monitored noisy oscillator}
\title{Quantum signatures in first-passage statistics of noisy monitored oscillators}
\title{Quantum-classical crossover in noisy monitored oscillators}

\author{Joseph M. Ryan}
\email{Contact author: joseph.ryan@duke.edu}
\author{Simon Gorbaty}

\author{Stephen W. Teitsworth}
\author{Crystal Noel}

\affiliation{Duke Quantum Center, Department of Electrical and Computer Engineering and Department of Physics, Duke University, Durham, NC 27708, USA}%

\date{\today}

\begin{abstract} 
The quantum first-passage problem involves stochastic trajectories conditioned on measurement outcomes.  The timing statistics of such trajectories remain largely unexplored in open quantum systems.  Here, we investigate the first-passage time to an energy threshold for a ubiquitous model: a harmonic oscillator driven by classical additive noise.  We find that projective measurements and energy quantization lead to substantial differences between the quantum and classical first-passage-time distributions at low thresholds, while these differences gradually diminish as the threshold energy increases.  We treat the problem using both ensemble-averaged conditioned density-matrix dynamics and trajectory-resolved stochastic pure-state dynamics.  The two descriptions yield indistinguishable timing statistics. Quantization effects appear in the ensemble-level phase-space distributions of the surviving states and vanish at larger threshold energies. Individual trajectories reveal emergent quantum signatures from the repeated measurements, such as persistent Wigner negativity.  Our results provide a framework for using first-passage processes to create measurement-induced nonclassical resource states and to study the quantum-classical crossover of monitored systems.

\end{abstract}

\maketitle
\section{Introduction}
\begin{figure}[h!]
    \centering
    \includegraphics[width=1\linewidth]{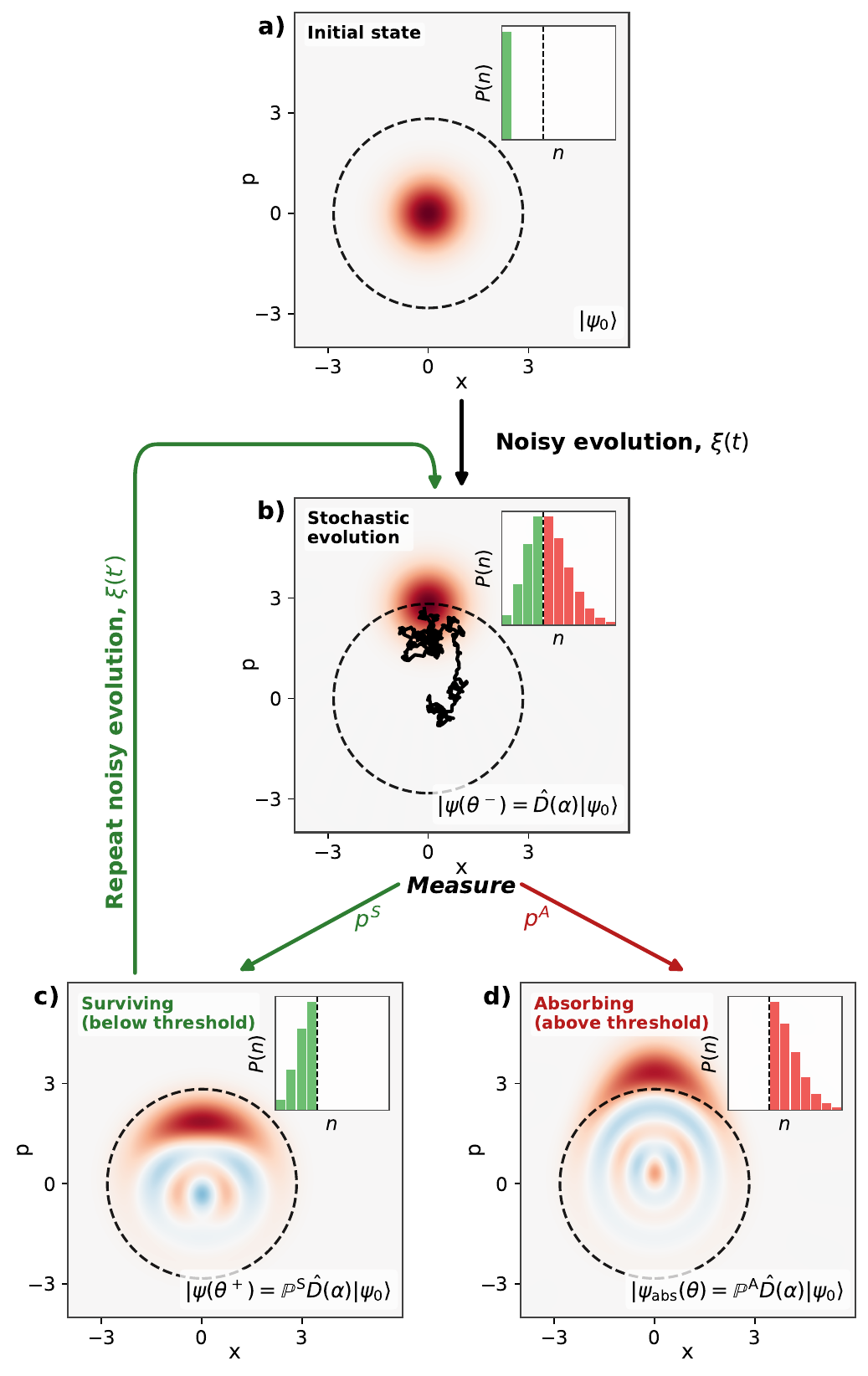}
    \caption{Illustration of a single trajectory of the first-passage process generated using a Monte Carlo wavefunction method. The Wigner function $W(x,p)$ of the wavefunction at each stage is shown. (a) Starting in the ground state $\ket{0}$, the noise $\xi(t)$ randomly displaces the wavefunction for a duration $\theta$, producing the state (b), after which time the energy-barrier measurement with threshold $N_B$ is performed (here with $N_B = 4$).  The measurement truncates the wavefunction, as shown in the histogram of occupation probabilities.  The two possible postmeasurement states (c) and (d) are shown.  The process repeats until an \textit{absorbing} outcome is obtained.}
    \label{fig:illustration}
\end{figure}
Noisy dynamical processes are generally described through ensemble observables, such as average motion, time-correlations, or probability densities, since individual trajectories fluctuate.  These observables characterize the collective behavior of many realizations, but they do not directly address properties of individual trajectories.  This is precisely the setting in which first-passage-time distributions become useful.  These distributions, which characterize the probability that a stochastic process first reaches a specified threshold or target at a given time, have been extensively studied~\cite{Redner_2001, Hanggi_RevModPhys.62.251, Teitsworth_PhysRevLett.109.026801, Teitsworth_EPJB_2019}, and have proven useful in many applications ranging from Schrödinger's clarification of Millikan's oil drop experiment~\cite{Schroedinger_FPT} to climate science~\cite{Climate_1, Climate_2, Climate_3}.  

The study of first-passage-time distributions in quantum systems is comparatively recent~\cite{quantum_walk_search_6, quantum_walk_search_7, Friedman_PRE, Kessler_PRL, Yin_Barkai, kewmingFPT}.
Quantum first-passage problems differ from classical ones because superposition and measurement backaction affect both the dynamics and the definition of the first-passage time.
For example, for closed quantum systems, sufficiently frequent measurements suppress escape events through the quantum Zeno effect~\cite{Misra_1976, Yin_Barkai}.  More generally, these questions are tied to the subtlety that time in quantum mechanics enters as a parameter rather than as a standard observable~\cite{das_quantum_time_Nature}.  Quantum first-passage statistics are also of growing interest in applications, ranging from quantum-walk search problems~\cite{quantum_walk_search_1, quantum_walk_search_2} to quantum metrology protocols based on sequential projective measurements~\cite{mentesoglu2026sequentialmeasurementsresourcequantum}.

Quantum first-passage problems admit a formulation as monitored quantum systems, in which repeated measurements generate stochastic dynamics conditioned on the measurement record~\cite{kewmingFPT}.  In this case, measurements both extract information and induce backaction on the system state~\cite{Wiseman_1996, Pinol_2024, Piper_2025}. Such conditioned dynamics have long been observed experimentally, for example, in cavity-QED experiments where probe atoms sequentially monitor the field and steer it stochastically toward photon-number states~\cite{Haroche_2007}. More recently, this trajectory-based viewpoint has found broad applicability in the context of quantum control and feedback, where the measurement record is used both to infer and to steer the system dynamics~\cite{wilson_2026, Lowen_2025, Pixley_2022, Wilson_2023, Landi_PRL, Landi_PRA_II, Landi_PRA}.

Here, we focus on the relatively unexplored problem of the first-passage-time statistics of a monitored open quantum system.  We investigate a model inspired by recent experimental measurements of the first-passage-time statistics of a monitored noisy oscillator to an energy threshold using quantum projective measurements~\cite{Ryan2025}, illustrated in Fig.~\ref{fig:illustration}.  We show that for low threshold energies, quantization effects produce substantial deviations between the quantum mechanical and the classical calculations of the first-passage-time statistics.  Furthermore, we demonstrate that repeated projective measurements associated with the first-passage process generate nonclassical quantum states, despite the presence of external noise driving the oscillator.  We show that operational observables such as first-passage times can become effectively classical before the conditioned quantum state dynamics do.  

We approach this problem using both a density matrix quantum master equation approach and a quantum trajectory-based approach.  While both yield indistinguishable first-passage-time statistics, the trajectory picture reveals the mechanisms by which the measurements generate nonclassical conditioned states despite the noise and more closely mirrors the nature of experimental realizations.  We present analytic calculations for the quantum first-passage-time distributions and contrast them with their classical counterparts.  We provide predictions throughout that can be tested, for example, in trapped-ion systems~\cite{Ryan2025}.  

\section{Model: the noisy harmonic oscillator}\label{section:model}

We consider a prototypical model: an undamped oscillator mode driven by weak additive Gaussian white noise, whose classical dynamics obey a Langevin equation
\begin{equation}\label{eq:classical_langevin}
    \Ddot{x} + \omega_0^2x = \xi(t),
\end{equation} where $x$ is the position, $\omega_0$ is the natural frequency, and $\xi(t)$ is a zero-mean delta-correlated force per unit mass with $\langle \xi(t)\xi(t')\rangle = 2D\delta(t-t')$.  Without loss of generality, we set $D=1$ for the rest of this manuscript, as this gives a heating rate of $\frac{\text{d}}{dt}\langle E(t) \rangle = D = 1$, where $E(t) = \dot x^2 / 2 + \omega_0^2x^2 / 2$.  This model describes, among others, the heating of trapped ions due to electric-field noise, which is a leading source of decoherence and loss of fidelity for trapped-ion quantum computers~\cite{Brownnutt, Wineland_Monroe_RevModPhys}.  

We explore the first-passage-time distribution of this oscillator with respect to an energy barrier $E_B=(N_B+1/2)$.  Throughout, we express energy in units of $\hbar\omega_0$.  Classically, the \textit{surviving} domain is then $\mathcal{D}_C = \{E: 0\leq E<E_B\}$.  In the quantum description, the \textit{surviving} domain is defined in terms of the energy eigenstates $\ket{n}$ of the unperturbed oscillator: $\mathcal{D}_Q =  \text{span}\{\ket{0},\ket{1},\ket{2}, ..., \ket{N_B-1}\}$.  The classical and quantum surviving domains differ at low energy as the quantum oscillator has a nonzero ground state energy $E=1 / 2$.  We assume that measurements to check whether escape has occurred are performed stroboscopically at intervals of duration $\theta$, and we investigate the $\theta\rightarrow 0$ limit of this stroboscopic projective measurement protocol.  The projectors associated with this measurement, which we call the energy-barrier measurement, are therefore $\mathbb{P}^\text{S} = \sum_{n=0}^{N_B-1} \ket{n}\bra{n}$ and $\mathbb{P}^\text{A} = \sum_{n=N_B}^\infty \ket{n}\bra{n}$, and correspond to \textit{survival} and \textit{absorption} (\textit{escape}),  respectively.  This measurement protocol has been implemented experimentally using quantum signal processing techniques in~\cite{Ryan2025}.   

The first-passage protocol is depicted in Fig.~\ref{fig:illustration}.  We first initialize the system in a chosen state, let it evolve under the noise for an interval $\theta$, and then perform the energy-barrier measurement.  In the quantum problem, first passage is defined operationally as the first time at which the \textit{absorbing} measurement outcome is obtained. By contrast, in the classical problem no distinction is made between crossing the threshold and observing that crossing.  The first-passage time is a random variable, whose distribution is known as the first-passage-time distribution.

\section{Ensemble-level dynamics}
We describe the coarse-grained ensemble-averaged dynamics using a density matrix $\rho = \sum_\psi P[\psi]\ket{\psi}\bra{\psi}$~\cite{Bruerer_and_pretrucione}.  The time-evolution master equation of a harmonic mode coupled to an infinite-temperature thermal amplitude reservoir, which is the quantum analogue of (\ref{eq:classical_langevin}) (provided that $\omega_0^{-1}\ll 1/\langle \dot n \rangle$), is given by~\cite{Turchette_engineered_reservoir}
\begin{equation}\label{eq:master_eq}
    \Dot{\rho}    
= \frac{\langle \dot n \rangle}{2} (2a\rho a^\dagger - a^\dagger a \rho - \rho a^\dagger a + 2a^\dagger \rho a -a a^\dagger \rho -\rho a a^\dagger),
\end{equation}
where $a^\dagger (a)$ are the usual creation (annihilation) operators for the oscillator.  We now use dimensionless time $t$ such that the heating rate is unity: $\langle \dot n \rangle = \frac{d}{dt}\Tr [a^\dagger a \rho(t)] = 1$.  Since we are considering the infinite-temperature bath, the dynamics generated by (\ref{eq:master_eq}) exhibit pure diffusion and indefinite heating. This master equation, which can be written as $\dot \rho = \mathcal{M}\rho$, is of the Lindblad form~\cite{Bruerer_and_pretrucione} and defines the propagator $\rho(t') = e^{\mathcal{M}(t'-t)}\rho(t)$.

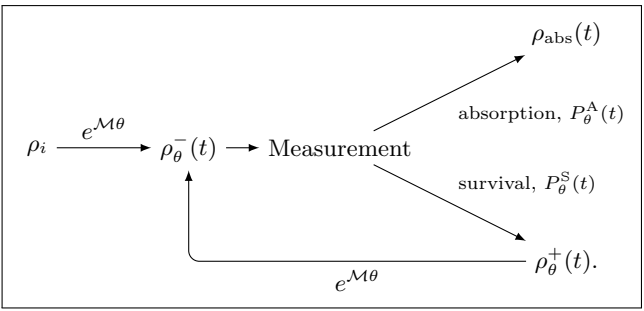
\begin{figure}[b]
\centering
\fbox{
\begin{tikzpicture}[baseline=(current bounding box.center),>=latex]
  \node (rho0) at (-2,0) {$\rho_i$};
  \node (rhom) at (0,0) {$\rho_\theta^-(t)$};
  \node (meas) at (2,0) {$\text{Measurement}$};
  \node (cont) at (5,-1.5) {$\rho_\theta^+(t).$};
  \node (abs)  at (5,+1.5) {$\rho_{\mathrm{abs}}(t)$};

  \draw[->] (rho0) -- node[above] {$e^{\mathcal{M}\theta}$} (rhom);
  \draw[->] (rhom) -- (meas);
  \draw[->] (meas) -- node[above right] {\scriptsize survival, $P_\theta^\text{S}(t)$} (cont);
  \draw[->] (meas) -- node[below right] {\scriptsize absorption, $P_\theta^\text{A}(t)$} (abs);

  \draw[<-, rounded corners] (rhom) |- node[pos=0.75, below] { $e^{\mathcal{M}\theta}$} (cont);
\end{tikzpicture}}
\caption{Schematic of the repeated measurement protocol in the ensemble picture.  The initial state $\rho_i$ evolves under the master equation propagator $e^{\mathcal{M}\theta}$ to the premeasurement state $\rho_\theta^-(t = \theta)$.  A measurement then produces either a \textit{survival} outcome, with probability $P_\theta^\text{S}(t=\theta)$, leaving the system in the postmeasurement state $\rho_\theta^+(t=\theta)$, or an \textit{absorption} outcome, with probability $P_\theta^\text{A}(t=\theta)$, leading to $\rho_\text{abs}(t=\theta)$.  In this case, the process stops.  Following a \textit{survival} outcome, the postmeasurement state is propagated for another interval $\theta$, and the cycle repeats.}
\label{fig:quantum_process_master_eq}
\end{figure}

The first-passage process (outlined schematically in Fig.~\ref{fig:quantum_process_master_eq}) consists of initializing the system in state $\rho_i$, followed by repeated intervals of noisy evolution of duration $\theta$, each followed by an energy-barrier measurement, until an \textit{absorption} outcome occurs.
We assume that the measurements themselves are fast relative to the measurement interval $\theta$.  In that case, the normalized density matrix representing the conditioned \textit{surviving} states immediately after the $m$-th measurement at $t_m=m\theta$ is given by 
\begin{equation}\label{eq:repeated_measurements}
    \rho^+_{\theta}\big(t_m\big) = \frac{(\mathcal{P}^\text{S}e^{\theta\mathcal{M}})^m [\rho_i]}{\Tr [(\mathcal{P}^\text{S}e^{\theta\mathcal{M}})^m [\rho_i]]}
\end{equation}
where $\mathcal{P}^\text{S}[\rho] := \mathbb{P}^\text{S}\rho{\mathbb{P}^\text{S}}^\dagger$.  These states $\rho^+_{\theta}(t_m)$ represent the ensemble of trajectories for which all measurements have resulted in \textit{survival} up to time $t_m$, and could in principle be reconstructed experimentally using standard quantum state tomography techniques~\cite{Wineland_Monroe_RevModPhys, Fluhmann2020}.  The subscript in $\rho_\theta^+(t)$ is used to emphasize the dependence of the surviving states on the measurement interval, as $\rho_\theta^+(t) \neq \rho_{\theta'}^+(t)$ for $\theta \neq \theta'$.  The corresponding discrete quantum first-passage-time distribution (FPTD) for a given interval $\theta$ is then given by
\begin{equation}\label{eq:def_FPTD}
P^\text{FPT}_{\theta}(t_m) = \prod_{k=1}^{m-1}P_\theta^{\text{S}}(t_k) \cdot P_\theta^{\text{A}}(t_m)    
\end{equation}
where $P_\theta^{\text{S}(\text{A})}(t_k)$ denotes the conditional probability of obtaining a \textit{survival} (\textit{absorption}) outcome from the $k^\text{th}$ measurement, given that all prior measurements resulted in \textit{survival}.   
\begin{figure}
    \centering
    \includegraphics[width=0.99\linewidth]{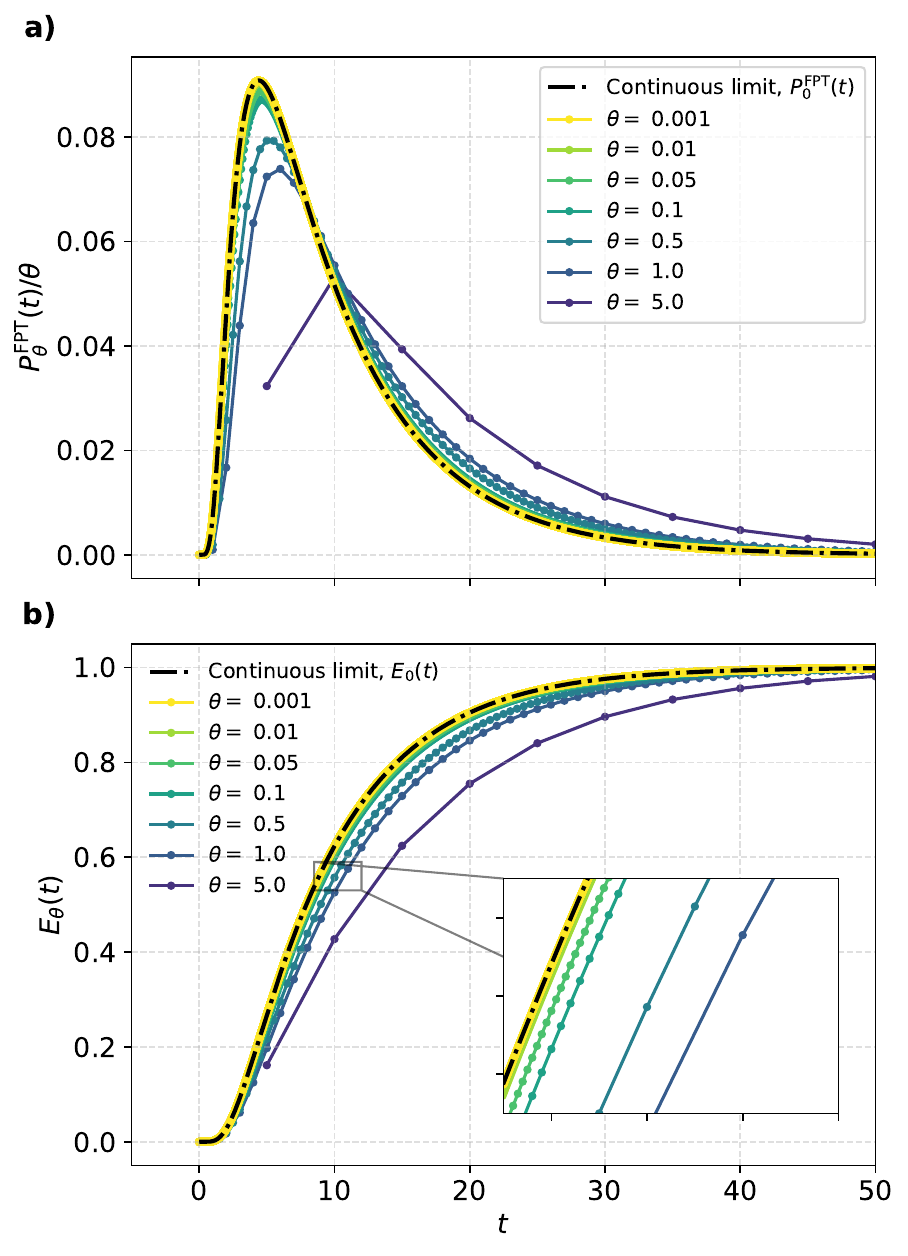}
    \caption{(a) First-passage-time distributions and (b) cumulative escape probability distributions for $N_B=10\, , N_0 = 0$ and various $\theta$.  For finite-$\theta$, these are discrete distributions defined on the grid $t=m\theta$, with $m=1,2,\ldots$.  We also show the infinitely frequent measurement limit densities $P^\text{FPT}_{0}(t)$ and $E_{0}(t)$.}
    \label{fig:escape_prob}
\end{figure}

We numerically evaluate (\ref{eq:repeated_measurements}) to obtain the discrete quantum first-passage distribution (\ref{eq:def_FPTD}) and the discrete cumulative escape probability $E_{\theta}(t)$ which is defined as
\begin{equation}
\begin{aligned}
E_{\theta}(t_m)&=\sum_{i=1}^{m} P^{\mathrm{FPT}}_{\theta}(t_i) \\
&= 1 - \Tr [(\mathcal{P^\text{S}}e^{\theta\mathcal{M}})^m [\rho_i]] \\
&= 1 - \prod_{k=1}^{m}P_\theta^\text{S}(k\theta).
\end{aligned}    
\label{eq:placeholding}
\end{equation}
Some examples of both the quantum first-passage-time distribution and the related cumulative escape distribution are shown in Fig.~\ref{fig:escape_prob}.  These are discrete distributions defined on the time grid $t_m=m\theta$.  In order to compare FPTDs corresponding to different measurement intervals $\theta$ on an equal footing without normalization artifacts, we plot $P_{\theta}^{\mathrm{FPT}}(t)/\theta$ in Fig.~\ref{fig:escape_prob}a, thereby removing trivial `bin-width' effects.  By contrast, the cumulative escape probability in Fig.~\ref{fig:escape_prob}b is not affected by this discretization dependence and therefore permits a more direct comparison across different values of $\theta$.  We observe that as $\theta\to0$, the discrete stroboscopic first-passage statistics converge smoothly to a continuous-time limit corresponding to infinitely frequent projective measurements.  We now explore this limit in more depth.

\subsection{Infinitely frequent measurement limit}

The quantum problem in the limit $\theta\to0$ becomes analytically tractable as the stroboscopic dynamics reduce to continuous-time dynamics, allowing us to apply continuous-time techniques and derive exact results.  In this limit, the discrete first-passage probabilities defined on the grid $t_m = m\theta$ converge to a continuous first-passage-time density according to
\begin{equation}
    P^\text{FPT}_{0}(t_m) = \lim_{\theta\rightarrow 0 } \frac{P^\text{FPT}_{\theta}(t_m)}{\theta},
\end{equation}
where $t$ is held fixed and $m=t/\theta\to\infty$.  To analyze this limit, we begin by considering the corresponding unnormalized conditional states $\tilde\rho_0(t)$ in the surviving subspace rather than working with the normalized conditioned surviving states in (\ref{eq:repeated_measurements}).  For clarity, we distinguish these from the normalized surviving states $\rho_{0}^+(t)$.  
We expand the stroboscopic propagator for small $\theta$,
\begin{equation}
    e^{\mathcal{M}\theta} = \mathbb{I} + \theta\mathcal{M} + \mathcal{O} (\theta^2),
\end{equation}
so that
\begin{equation}
\mathcal{P}^\text{S} e^{\mathcal{M}\theta} = \mathcal{P}^\text{S} + \theta \mathcal{M}^{(N_B)} + \mathcal{O} (\theta^2),
\end{equation}
where $\mathcal{M}^{(N_B)}:=\mathcal{P}^\text{S}\mathcal{M}\mathcal{P}^\text{S}$ is the truncated superoperator obtained by restricting $\mathcal{M}$ to the surviving subspace.  Taking the limit of $t/\theta$ repeated applications as $\theta\rightarrow 0$ then yields the continuous-time evolution generated by $\mathcal{M}^{(N_B)}$.
We find that the unnormalized states satisfy
\begin{equation}
\begin{aligned}
    \tilde\rho_0(t)&=\lim_{\theta\to 0}(\mathcal{P}^{\mathrm S}e^{\theta\mathcal M})^{t/\theta}[\rho_i] \\
    &= e^{\mathcal{M}^{(N_B)}t}[\rho_i].
\end{aligned}
\label{eq:unnormalized_surviving_states}
\end{equation}

\begin{figure}
    \centering
    \includegraphics[width=1\linewidth]{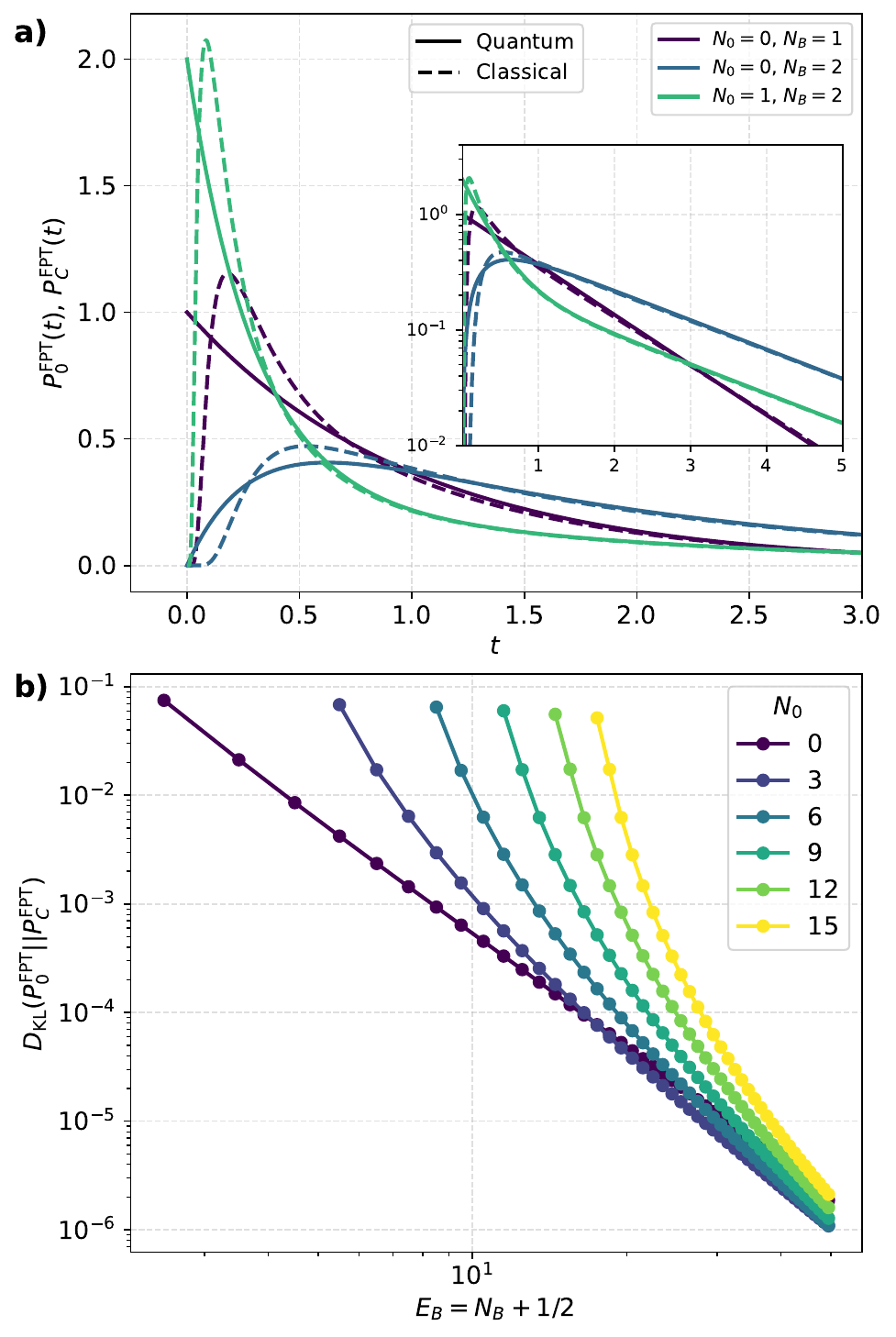}
    \caption{(a) Continuous-time quantum (solid lines) and classical (dashed lines) FPTDs for a given starting energy $E_0 = (N_0+1/2)$ and threshold energy $E_B = (N_B+1/2)$.  (b) Kullback–Leibler divergence  $D_\text{KL}(P_0^\text{FPT}||P_C^\text{FPT})$ between quantum and classical FPTDs for different initial and threshold energies.  The points corresponding to $N_B-N_0 = 1$ are omitted as the classical density essentially vanishes at zero time, and so the KL integrand is logarithmically divergent.  }
    \label{fig:fptds_q_vs_c}
\end{figure}

The trace of the unnormalized surviving state $\tilde\rho_0(t)$ gives the survival probability $\Tr [\tilde\rho_0(t)] = S_{0}(t) = 1 - E_{0}(t)$.  In the limit $\theta\to\ 0$, the dynamics reduces to an effective evolution on the surviving subspace $n<N_B$, with the threshold $N_B$ acting as an absorbing boundary. Writing $P_n(t) = \bra{n}\tilde\rho_0(t)\ket{n}$, the continuous-time population dynamics below the threshold obey
\begin{equation}\label{eq:master_equation_Birth_death_main_text}
\begin{split}
    \partial_tP_n(t) &= (n+1)P_{n+1}(t) - [2n+1]P_n(t) +nP_{n-1}(t)
\end{split}
\end{equation}
with an absorbing boundary condition $P_{N_B}(t) = 0$.   These equations define a finite-dimensional truncated generator on the surviving subspace.  The associated eigenvalue equation satisfies a three-term recursion identical to that of the Laguerre polynomials~\cite{arfken2013mathematical}.  We use this recurrence to derive the corresponding first-passage density $P^\text{FPT}_{0}$ in the infinitely frequent measurement limit with Laplace transforms.  We obtain, for an initial Fock state $\rho_i=\ket{N_0}\bra{N_0}$ with $N_0<N_B$,
\begin{equation} \label{eq:exact_result}
    P^\text{FPT}_{0}(t| N_0, N_B) = \sum_i\frac{\mathcal{L}_{N_0}(\lambda_i^{N_B})}{\mathcal{L}_{N_B-1}^{(1)}(\lambda_i^{N_B})}e^{-\lambda_i^{N_B}t}
\end{equation}
where $\mathcal{L}_m(x)$ is the $m$-th Laguerre polynomial and $\lambda_i^{m}>0$ are its $m$ distinct and positive roots labeled in increasing order as $\lambda_1^{m}, \lambda_2^{m}, ..., \lambda_m^{m}$.  See Appendix~\ref{app:QFPTD} for the derivation and consideration of more complex initial conditions.  The FPTD is a finite sum of exponential decay modes whose rates are the roots of the Laguerre polynomial $\mathcal{L}_{N_B}$.  Some examples of $P^\text{FPT}_{0}$ are shown in Fig.~\ref{fig:fptds_q_vs_c}a.

We find that for $N_B=1$, the FPTD is a pure exponential.  This is expected since each measurement that does not result in \textit{escape} projects the state back into the ground state, and thus there is a fixed escape probability at each measurement.  (For finite nonzero $\theta$, the same logic applies and the FPTD is a geometric distribution with probability $p = \theta / (\theta+1)$.)  More generally, for the cases where $N_B-N_0 = 1$ there is no initial ballistic part, but the decay is multiexponential.  For example, for $N_0 = 1, N_B = 2$ the FPTD is a biexponential, as is clear from the log-scaled inset in Fig.~\ref{fig:fptds_q_vs_c}a.  These effects occur as a result of the quantization of energy levels.

We compare the infinitely frequent measurement limit quantum first-passage density $P_{0}^\text{FPT}(t)$ in (\ref{eq:exact_result}) with the corresponding classical continuous-time first-passage density $P_C^\text{FPT}(t)$, which we derive directly from (\ref{eq:classical_langevin}) using classical stochastic averaging in Appendix~\ref{app:classical_problem}.  We note that the stochastic-averaging approximation is valid when the heating per oscillation is small, which is analogous to the weak-coupling assumption used in deriving the quantum master equation.  The comparison is shown in Fig.~\ref{fig:fptds_q_vs_c}a.  We quantify the distance between these continuous-time FPTDs using the Kullback–Leibler (KL) divergence~\cite{Kullback_1951, Raghu_2025}, which is defined as
\begin{equation}\label{eq:KL_divergence_definition}
    D_\text{KL}(P_0^\text{FPT}||P_C^\text{FPT}) = \int_0^\infty dt P_{0}^\text{FPT}(t) \log \frac{P_{0}^\text{FPT}(t)}{P_C^\text{FPT}(t)},
\end{equation}
and is shown in Fig.~\ref{fig:fptds_q_vs_c}b.  The short- and intermediate-time parts of the FPTDs retain the strongest signatures of the discrete level structure as shown in Fig.~\ref{fig:fptds_q_vs_c}.  This is particularly apparent when the threshold $N_B$ is close to the initial level $N_0$, where the escape process is governed by fewer discrete modes. As $N_B-N_0$ increases, more levels participate before absorption and the quantum first-passage statistics become progressively coarse-grained and approach the classical result.  The decreasing KL divergence signals the crossover from a distinctly quantum regime into a more classical regime.  The long-time parts of the two distributions are comparatively less distinguishable, since both are dominated by a single effective decay scale.

\subsubsection{Long-time quasi-stationary surviving states}

At long times, the infinitely frequent measurement limit quantum first-passage density decays exponentially as $\sim e^{-\lambda_1^{N_B}t}$ where $\lambda_1^{N_B}$ is the smallest root of the Laguerre polynomial $\mathcal{L}_{N_B}$.  We note that this smallest root scales inversely with the threshold energy $E_B$ and is approximately $\lambda_1^{N_B} \approx 2.404^2 / (4E_B) \propto 1/E_B$ (see Appendix~\ref{app:classical_problem} for details).  In this regime, the unnormalized surviving state is dominated by the slowest decaying eigenmode of $\mathcal{M}^{(N_B)}$, 
\begin{equation}
    \tilde\rho_0(t)\sim e^{-\lambda_1^{N_B}t} \rho_0^\text{qs},\hspace{15pt} t\to\infty,
\end{equation}
so that the normalized surviving state $\rho_0^+(t) = \tilde\rho_0(t) / \Tr[\tilde\rho_0(t)]$ converges to the quasi-stationary state $\rho_0^\text{qs} = \lim_{t\to\infty}\big[ \rho^+_0(t) \big]$.  This state is diagonal in the Fock basis, with occupation probabilities
\begin{equation}\label{eq:steady_state}
    [\rho_0^\text{qs}]_{nn} = \frac{1}{C} \mathcal{L}_n(\lambda^{N_B}_1),
\end{equation}
where $C = \mathcal{L}_{N_B}^{(1)}(\lambda_1^{N_B})$ is a normalization factor.  See Appendix~\ref{app:surviving_states} for the derivation.  The time evolution of the conditioned surviving states in the infinitely frequent measurement limit may be found in an analogous way.  

Physically, we see that the conditioning on survival outcomes preferentially selects realizations whose occupation remains far from the absorbing threshold, thereby producing distinctly nonthermal states.  This quasi-stationary state serves as a useful reference state for subsequent analysis of measurement-induced nonclassicality and the quantum-classical crossover in phase space.

\subsection{Ensemble-level nonclassicality of the surviving states}

Although the master equation describing the evolution between measurements (\ref{eq:master_eq}) can be mapped onto a classical Markov (birth-death) process~\cite{Gardiner_quantum_noise} and represented in phase space by a Fokker-Planck-type description, this does not imply that the measurement-conditioned dynamics admit a classical interpretation.  In our case, the repeated projective threshold measurements alter the evolution through the conditioning in (\ref{eq:repeated_measurements}), and the measurement-conditioned surviving states are not constrained to remain classical.  

\begin{figure}
    \centering
    \includegraphics[width=1\linewidth]{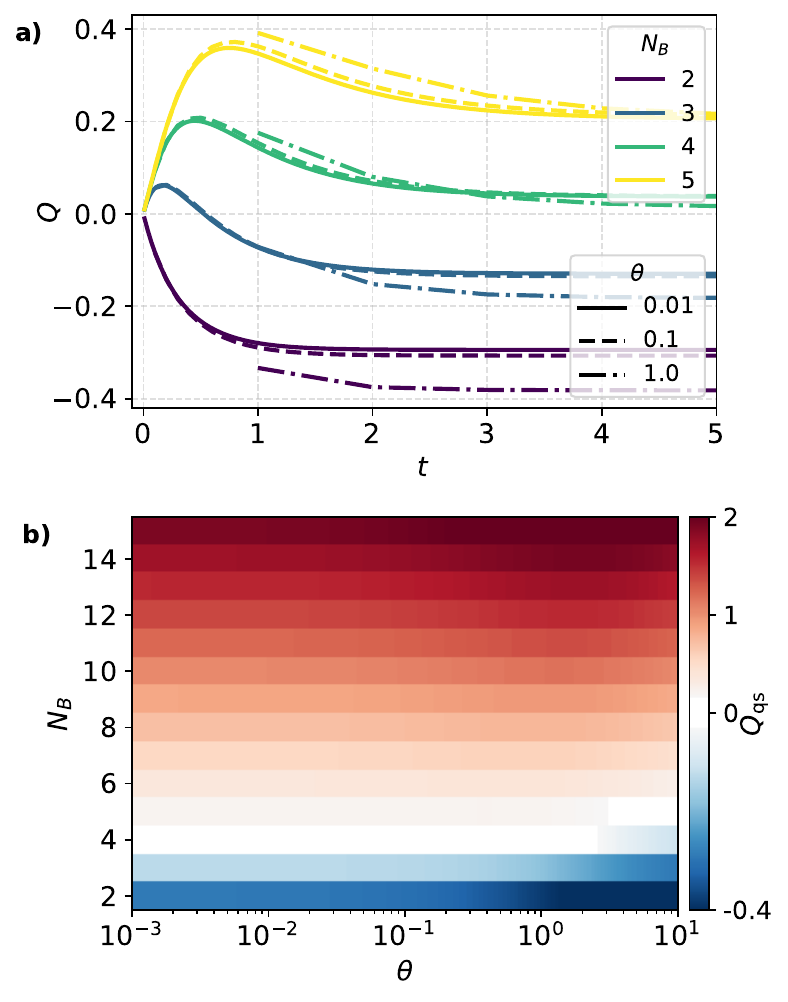}
    \caption{(a) Time evolution of the $Q$ parameter of surviving states for different $N_B$ and $\theta$.  (b) Mandel-$Q$ parameter of the quasi-stationary motional state $\rho_\theta^\text{qs}$ of the first-passage process for different values of $N_B$ and $\theta$.}
    \label{fig:mandel_Q}
\end{figure}

More precisely, a quantum state is considered nonclassical if its $P$-representation is not a classical probability density~\cite{vogel_non_classical, Richter2002}.  Due to the measurements considered in this manuscript, every surviving state after a threshold measurement has support only on the finite-dimensional surviving subspace. Except for the vacuum state $\ket{0}$, any such finite-support state is nonclassical, since no nonnegative $P$-representation can reproduce a density operator with compact support in the Fock basis~\cite{non_gaussianity_detection_Park}. Thus, the nonclassicality of the surviving ensemble is a direct consequence of the threshold-conditioning protocol itself.  While mathematically rigorous, this conclusion is also a rather specific consequence of the ideal threshold-conditioning protocol. We therefore also examine below more physically transparent signatures of nonclassicality.

\subsubsection{Mandel $Q$-parameter}
As an experimentally accessible witness of this nonclassicality, we consider the Mandel $Q$-parameter~\cite{Mandel_Wolf_1995}
\begin{equation}
    Q = \frac{\langle \hat n^2\rangle - \langle\hat n\rangle^2}{\langle\hat n\rangle} - 1,    
\end{equation}
where $\hat n = \hat a^\dagger \hat a$ is the usual number operator.  Negative values of $Q$ correspond to sub-Poissonian phonon number statistics, a widely accepted indicator of nonclassicality~\cite{Mandel_Wolf_1995}.  These negative values imply that the $P$-representation cannot be interpreted as a probability density.  We show the time evolution of the $Q$ parameter of the surviving states for selected $N_B$ in Fig.~\ref{fig:mandel_Q}a.   At short times, the $Q$ parameter increases linearly, which is consistent with thermal state evolution.  (Except for $N_B=2$, where $Q$ is always negative for every nonvacuum state supported only on $\{{|0\rangle,|1\rangle}\}$, as $\langle \hat n^2\rangle=\langle \hat n\rangle$ and hence $Q=-\langle \hat n\rangle<0$, irrespective of any coherence between $\ket{0}$ and $\ket{1}$.)  However, once the occupation near the threshold becomes appreciable, the subsequent evolution is strongly modified by the threshold conditioning and the $Q$ parameter decreases and subsequently approaches a constant.  We show the $Q$ parameter of the long-time quasi-stationary state $\rho_\theta^\text{qs} = \bigg[\lim_{t\to\infty}\big[ \rho^+_\theta(t) \big]\bigg]$ in Fig.~\ref{fig:mandel_Q}b.  We find these states numerically and then calculate the $Q$ parameter.  In the limit $\theta\to 0$, the numerical results converge exactly to the analytical infinitely frequent measurement-limit solution given in~(\ref{eq:steady_state}).  We find that, over a wide range of $\theta$, $Q<0$ for $N_B\in\{2,3\}$.  (For $N_B=1$ the surviving state is $\ket{0}$ which is conventionally assigned the Poissonian value of $Q=0$.)    We emphasize that $Q\geq 0$ is necessary but not sufficient to guarantee classicality.  For example, squeezed-vacuum states can have $Q\geq 0$~\cite{walls2008}.

 \begin{figure}
    \centering
    \includegraphics[width=1\linewidth]{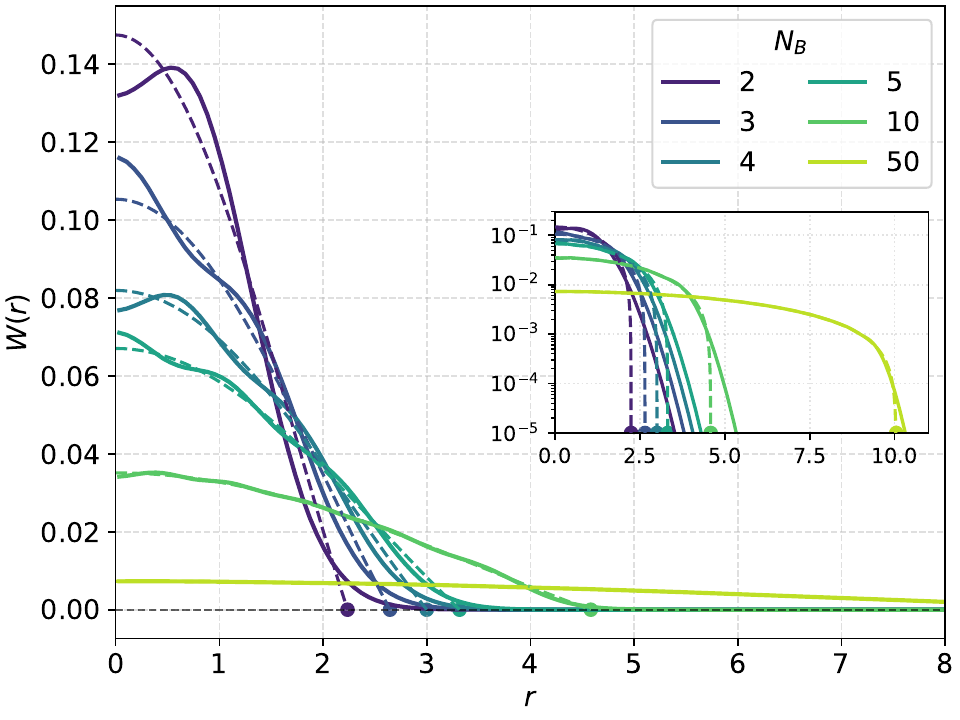}
    \caption{The radial profiles of the rotationally symmetric Wigner function $W(r = \sqrt{x^2 + p^2})$ of the quasi-stationary motional state (\ref{eq:steady_state}) are shown with solid lines.  The corresponding classical phase-space density is shown with dashed lines.  Colored dots are used to mark the energy barrier $r_B = \sqrt{2E_B}$.}
    \label{fig:wigner}
\end{figure}

\subsubsection{Wigner function}\label{section:wigner_ensemble}
A complementary characterization of the measurement-conditioned surviving ensemble is provided by its Wigner function.  For the dynamics considered here, the measurement-conditioned surviving states are mixed states that are diagonal in the Fock basis (provided that their initial states $\rho_i$ were diagonal too), so their Wigner functions are rotationally symmetric,
\begin{equation}
    W(x,p) = W(r), \hspace{5pt} r^2 = x^2 + p^2.
\end{equation}
Therefore, for a diagonal surviving state $\rho^+(t)$, the Wigner function is given by
\begin{equation}
    W(r, t) = \frac{1}{\pi}e^{-r^2}\sum_{n=0}^{N_B-1}(-1)^n[\rho^+(t)]_{nn}\mathcal{L}_n(2r^2).
\end{equation}

As an illustrative example, Fig.~\ref{fig:wigner} shows the Wigner radial profiles of the continuous-measurement quasi-stationary states given by (\ref{eq:steady_state}).  These profiles display radial oscillations in $r$ that are most pronounced for small $N_B$ and appear to progressively diminish as $N_B$ increases, showing that these states are non-Gaussian~\cite{Walschaers2021}.  However, non-Gaussianity alone is not a uniquely quantum feature of the first-passage ensemble.  The quantum signature is instead contained in the detailed oscillatory radial structure, which is absent from the corresponding classical phase-space density (\ref{eq:classical_phase_space_density}) derived in Appendix~\ref{app:classical_problem} and plotted with dashed lines in Fig.~\ref{fig:wigner}.

Since the surviving conditional states are diagonal in the Fock basis, these oscillations do not arise from phase coherence between different number states. Rather, they originate from the oscillatory Laguerre kernels associated with the discrete set of occupied Fock levels. The effect is most pronounced for small $N_B$ as in this deeply quantized regime the phase-space dynamics resolve the granularity of the oscillator spectrum, and this discrete spectral structure leaves a visible imprint in phase space. As $N_B$ increases, the quasi-stationary state receives contributions from more levels, the spectrum becomes effectively coarse-grained, and the radial oscillations are progressively washed out as shown in Fig.~\ref{fig:wigner}. The disappearance of the oscillations at large $N_B$ thus signals the crossover toward a more classical, continuous-energy description.  There is another small difference between the quantum and the classical phase-space structure.  In the classical case, due to the energy thresholding, there is no occupation probability above the barrier radius $r_B = \sqrt{2E_B}$.  In the quantum case, by contrast, there is no analogous hard cutoff as the Wigner function extends over all r.  This difference is also shown in Fig.~\ref{fig:wigner}.

Although the ensemble quasi-stationary Wigner functions shown in Fig.~\ref{fig:wigner} are nonnegative, their oscillatory structure shows that the conditioned long-time state is not simply a classical truncated low-energy distribution. We note that the nonnegativity holds at all times (not just in the long-time limit) for any passive initial state (one with non-increasing Fock populations, such as the ground state or a thermal state) as the conditioned evolution (\ref{eq:master_equation_Birth_death_main_text}) preserves passivity (see Appendix~\ref{app:surviving_states}), and passive states have nonnegative Wigner functions~\cite{Herstraeten_2021}. Below, we consider a trajectory-resolved description of the measurement-conditioned surviving states and see that at the level of individual trajectories the Wigner functions of these states contain negativity that is averaged out in the ensemble picture.  We discuss the non-Gaussianity of the ensemble Wigner functions further in Appendix~\ref{app:nonGaussianity}.

\section{Trajectory-resolved description}
While the first-passage timing statistics can be obtained from the ensemble-level description as shown above, there are aspects of the measurement-conditioned surviving states that cannot.  This is because the density matrix representing the surviving conditioned states $\rho^+(t)$ is an ensemble average over stochastic noise realizations and measurement outcomes, and therefore averages out features tied to individual conditioned evolutions.  Ensemble averaging therefore discards trajectory-resolved information, including fluctuations between individual realizations and nonlinear quantities evaluated along surviving trajectories, such as the Wigner negativity~\cite{Bruerer_and_pretrucione, Bruerer_Petrucione_variances}.  Figure~\ref{fig:illustration} illustrates this point by showing the Wigner function for a single representative noisy realization that begins in the ground state $\ket{0}$, where nonclassical features appear directly as regions of negativity, whereas in the ensemble-averaged case there are no negative regions as discussed in Sec.~\ref{section:wigner_ensemble} and shown in Fig.~\ref{fig:wigner}.

\begin{figure}
    \centering
    \includegraphics[width=1\linewidth]{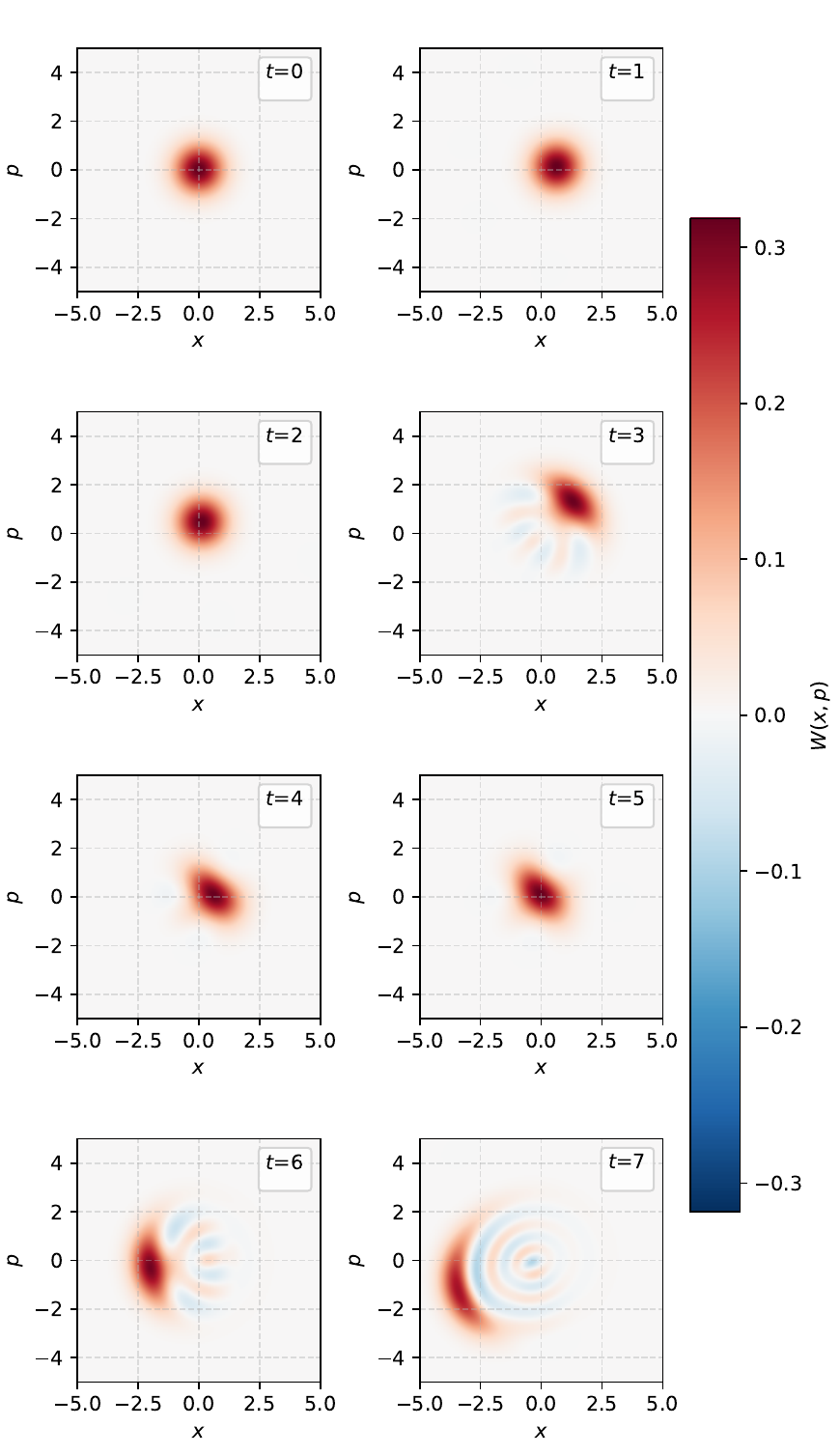}
    \caption{Time evolution of the Wigner function $W(x,p)$ for a sample trajectory beginning in the ground state $\ket{0}$, with $\theta = 1$ and $N_B=5$.  This trajectory has a first-passage time $T=7$.  The Wigner function is evaluated after each measurement.  Between measurements, the wavefunction is stochastically displaced under (\ref{eq:displaced}), so the Wigner function is shifted without distortion.  Note that the Wigner function is negative at some times, indicating nonclassicality.  }
    \label{fig:wigner_trajectory_time_evolution}
\end{figure}

The trajectory-resolved first-passage process proceeds as follows.  Each trial starts with the system initialized in the state $ \ket{\psi_0}$.   
The state then evolves under the influence of noise $\xi(t)$ which causes an initial wavefunction $\ket{\psi(t)}$ to evolve stochastically in the rotating frame according to~\cite{James_alpha_kicks_1998, richerme_2024measurementinducedheatingtrappedions}
\begin{equation}\label{eq:displaced}
        \ket{\psi(t)} \rightarrow \ket{\psi(t+\theta)} = \mathcal{D}(\alpha) \ket{\psi(t)},
    \end{equation}
    where
    \begin{equation} \label{eq:alpha}
        \alpha = i\sqrt{\frac{m}{2\hbar\omega_0}}\int_t^{t+\theta}dt' \xi(t')e^{i\omega_0 t'}
    \end{equation}
and where $\mathcal{D}(\alpha) = \exp\,(\alpha a^\dagger -\alpha^* a)$ is the displacement operator.  For delta-correlated noise, as is the case here, the ensemble average of (\ref{eq:displaced}) and (\ref{eq:alpha}) is equivalent to (\ref{eq:master_eq}).  The displacement $\alpha$ is randomly distributed according to a complex normal distribution, $\alpha\sim\mathcal{N_C}(0,\theta)$ such that $\langle|\alpha|^2\rangle = \theta$ with the noise normalization of Sec.~\ref{section:model} (unit mass, $D=1$ and dimensionless time such that $\langle\dot n\rangle = mD/\hbar\omega_0 = 1$).  This is expected as displacing the ground state by $\alpha$ adds $\langle n\rangle = |\alpha|^2$ quanta, so we require $\langle |\alpha|^2\rangle  = \langle \dot n\rangle\theta = \theta$.  In the absence of measurements, an initially coherent state is expected to remain coherent under the influence of the noise which performs random displacements.  

Before the first measurement the wavefunction is
\begin{equation}\label{eq:traje_1}
    \ket{\psi(\theta^-)} = \mathcal{D}(\alpha)\ket{\psi_0}.
\end{equation}
The energy-barrier measurement is then performed, and a \textit{survival} outcome is obtained with probability $p^\text{S}(\theta) = \bra{\psi(\theta^-)}\mathbb{P}^\text{S}\ket{\psi(\theta^-)}$.  The wavefunction after the measurement is then
\begin{equation}\label{eq:traje_2}
    \ket{\psi(\theta^+)} = \frac{ \mathbb{P}^\text{S}\ket{\psi(\theta^-)}}{\sqrt{p^\text{S}(\theta)}}.
\end{equation}
This sequence of interleaved evolutions and measurements continues until an \textit{absorption} measurement outcome is obtained, as illustrated in Fig.~\ref{fig:illustration}.

This trajectory picture also makes effects associated with the quantum nature of the measurement particularly transparent.  First, within a single trajectory, any projective measurement that yields \textit{survival} lowers or leaves unchanged the post-measurement energy value through measurement backaction by removing any above-threshold support.  Second, a trajectory whose mean energy is far below the threshold may still be terminated probabilistically if it contains populations in energy levels greater than the threshold.  Conversely, a trajectory whose mean energy is far above the threshold may not be terminated by the projective measurement.  These effects have no classical analogue.

We simulate the quantum first-passage process using a Monte Carlo wavefunction simulation~\cite{Molmer_1993} to sample $P[\psi(t)]$ and obtain the first-passage-time distributions.  Note that $P[\psi(t)]$ is a distribution over wavefunctions that have survived until at least time $t$.  The obtained FPTDs are indistinguishable (see Appendix~\ref{app:MonteCarlo}) to those obtained by direct numerical integration of (\ref{eq:repeated_measurements}) and (\ref{eq:def_FPTD}).  

\subsubsection{Wigner function negativity}
Beyond reproducing the ensemble-level first-passage-time statistics, a particularly useful application of the trajectory description is to follow the Wigner function along individual realizations, revealing nonclassical phase-space structure that is not obtainable by ensemble averaging.  For a pure state $\ket{\phi}$, the Wigner function is given by~\cite{Gardiner_quantum_noise}
\begin{equation}\label{eq:wigner_def}
     W_\phi( x,  p) = \frac{1}{\pi}\int dy\phi^*( x-y)\phi( x + y)e^{-2iy p}.
\end{equation}
We show the time evolution of the Wigner function of a single example trajectory in Fig.~\ref{fig:wigner_trajectory_time_evolution} which displays negativity, a nonclassical feature signaling the absence of a classical joint phase-space probability distribution for the noncommuting quadratures $x$ and $p$~\cite{Yoshikawa_2013}.  While displacements due to the noise simply translate the Wigner function without distortion, the projective measurement substantially modifies it, as also shown in Fig.~\ref{fig:illustration}.  In the trajectories picture, negativity is a natural consequence of the measurement as it truncates the Fock support of a displaced coherent-like state, and any pure state with bounded Fock support (other than vacuum) is non-Gaussian and hence Wigner-negative by Hudson's theorem~\cite{Hudson_1974}. 
\begin{figure}
    \centering
    \includegraphics[width=1\linewidth]{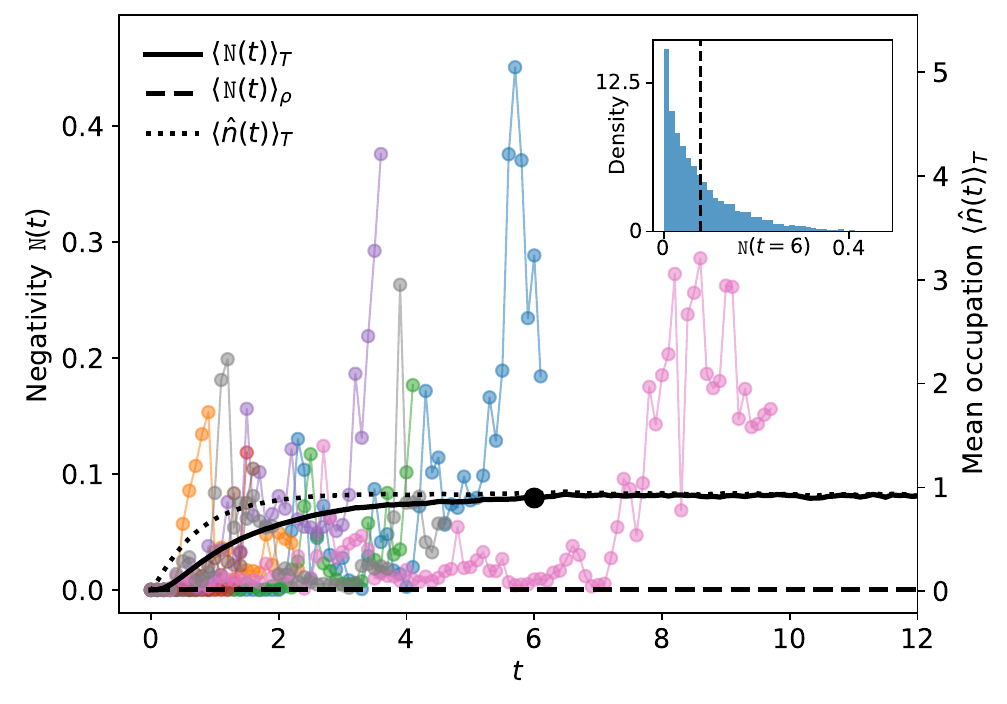}
    \caption{Negativity $\mathtt{N}(t)$ for some representative trajectories are shown in color.  We show the trajectory-averaged negativity $\langle \mathtt{N}(t)\rangle_T$ as a solid black line, whereas we show the averaged ensemble negativity $\langle \mathtt{N}(t)\rangle_\rho$ (which vanishes for the passive initial state $\ket{0}$; see Appendix~\ref{app:surviving_states}) as a dashed line.  The mean occupation of the surviving states $\langle \hat n(t)\rangle_T$ is shown as a dotted black line, and is referenced to the axis on the right.  The negativity and energy y-axes, on the left and right respectively, are scaled so that the steady-state mean energy and the steady-state trajectory-averaged negativity overlap to facilitate comparison of the timescales.  In the inset we show the distribution (which appears exponential) of simulated negativities at $t=6$.  For these simulated data, $\theta = 0.1$ and $N_B = 4$.  We simulate $5\times 10^4$ trajectories to estimate $P[\psi(t)]$.  Error bars on the computed means are too small to be seen.}
    \label{fig:wigner_trajectory_negativity}
\end{figure}

We quantify nonclassicality using the negativity (i.e., negative volume) of the surviving trajectories.  For a (possibly mixed) state $\rho$ the negativity $\mathtt{N}$ is defined as~\cite{negativity_kenfack_2004}
\begin{equation}
\begin{aligned}
    \mathtt{N}(\rho) &= \frac{1}{2} \bigg( \int\int dx\,dp \; \Big[\big|W_{\rho}(x,p)\big| - W_{\rho}(x,p)\Big]\bigg) \\
    &= \frac{1}{2}\bigg(\int\int dx\,dp  \; \big|W_{\rho}(x,p)\big| - 1\bigg),
\end{aligned}    
\end{equation}
where $W_{\rho}(x,p)$ is the Wigner function associated with the state $\rho$.  The Wigner function satisfies the normalization condition $\int d\vec\beta W(\vec \beta) = 1$.  

For an ensemble of pure state trajectories $\rho(t) = \mathbb{E}(\ket{\psi(t)}\bra{\psi(t)}) = \int D\psi D\psi^* P[\psi(t)]\ket{\psi(t)}\bra{\psi(t)}$, we may define the negativity in two ways.  We can define the trajectory-averaged negativity $\langle \mathtt{N}(t)\rangle_T$~\cite{Eastman2017}
\begin{equation}
    \langle\mathtt{N}(t)\rangle_T = \int D\psi D\psi^* P[\psi(t)]\mathtt{N}(\ket{\psi(t)}\bra{\psi(t)}).  
\end{equation}
Alternatively, we can define the negativity of the averaged ensemble $ \langle\mathtt{N}(t)\rangle_\rho$
\begin{equation}
    \langle\mathtt{N}(t)\rangle_\rho = \mathtt{N}\bigg(\int D\psi D\psi^* P[\psi(t)]\ket{\psi(t)}\bra{\psi(t)}\bigg).  
\end{equation}
The trajectory-averaged negativity can only be obtained in the trajectories picture, unlike the averaged ensemble negativity which can be obtained from both the trajectories picture and the ensemble density matrix picture.  This is because negativity is nonlinear in the states, and thus it cannot be obtained from the ensemble-averaged density matrix $\rho$ alone.  As discussed in Sec.~\ref{section:wigner_ensemble} and Appendix~\ref{app:surviving_states}, for passive initial states such as the ground state $\ket{0}$ considered here, the ensemble-averaged Wigner function is nonnegative at all times, so $\langle\mathtt{N}(t)\rangle_\rho = 0$.

In order to illustrate this difference, we show the time evolution of the negativity of a few sample pure-state trajectories, as well as the trajectory-averaged negativity in Fig.~\ref{fig:wigner_trajectory_negativity}.  The trajectory-averaged negativity increases from zero with time, until the quasi-stationary regime is reached. The non-zero long-time negativity shows that the energy-barrier measurement protocol causes the surviving states to develop nonclassical features that persist in time.  We note that in the absence of measurements, because the noise simply performs random displacements, the trajectory-averaged negativity is zero.  In order to make the contrast between the two negativities explicit, we also show the negativity of the averaged ensemble, which is zero, in Fig.~\ref{fig:wigner_trajectory_negativity}.  We also show the mean energy of the surviving states $\langle \hat n (t)\rangle_T$, which unlike the negativity can be obtained from the ensemble description.  We include it to compare the timescale at which the trajectory averaged negativity and the mean energy evolve.  Whereas the mean energy increases linearly from the initial time, the negativity increases more slowly.  Similarly, it appears that the negativity takes longer to achieve a steady state than the mean energy, which suggests that even though the averaged ensemble may have reached a quasi-stationary state, there are other trajectory-based observables that do not reach a steady-state at the same time.

The master-equation dynamics can be unraveled into stochastic pure-state trajectories, where quantum statistical ensembles are represented as a probability distribution (convex linear combination) of pure states $P[\psi(t)]$ on projective Hilbert space.  Whereas the trajectory ensemble uniquely determines the density matrix (and therefore all observables that are functions of the ensemble-averaged state), the converse is not true.  Many distinct trajectory ensembles can correspond to the same density matrix.  The diffusive/random-displacement unraveling we are using is not the only valid unraveling of the master equation (\ref{eq:master_eq}), as a jump unraveling where there are stochastic phonon-gain and phonon-loss jumps also gives the same ensemble behavior. However, both of these unravelings will yield different trajectory ensembles.  The trajectory-level negativity depends on the choice of unraveling as $\langle \mathtt{N}(t)\rangle_T$ is not a function of the ensemble-averaged state. Here, however, the diffusive unraveling is physically privileged as the noise is a classical force (\ref{eq:classical_langevin}) whose realization is, in principle, an accessible classical record, and the survival outcomes are recorded measurements. The conditioned pure states are therefore the correct predictive states for an observer with this record.  Although the trajectories considered here are generated by uncontrolled ambient noise, the same ensemble can be realized with programmed random displacements, making the negativity directly measurable through postselection on the measurement record and conditional state tomography~\cite{Pinol_2024, Fluhmann2020, Wineland_Monroe_RevModPhys}.

More broadly, this trajectory-level analysis serves as an example that quantum first-passage processes whose timing statistics approach those of their classical counterparts can still exhibit nonclassical conditioned dynamics.  Whether the process appears classical depends on the level of description (timing statistics, ensemble state, or individual trajectory), and not on the process itself.

\section{Discussion}
In this work, we have described the quantum first-passage problem of an experimentally ubiquitous model: an open noisy quantum system subject to quantum projective stroboscopic measurements.  We have shown that quantization substantially modifies the timing statistics when the initial and threshold energies are both of order $\sim\hbar\omega_0$.  These differences vanish as the relevant energies become large compared with the level spacing.  At the same time, this apparent emergent classicality of the first-passage statistics does not imply classical conditioned dynamics: the repeated measurements prepare surviving states with clear nonclassical signatures such as sub-Poissonian number statistics, nonpositivity of the $P$-representation, and trajectory-level Wigner negativity.  

Altogether, these results show that operational observables can appear classical before the underlying conditioned state dynamics do.  More broadly, our work identifies monitored first-passage problems as a useful setting for studying the emergence of classicality in open quantum systems, and for isolating the distinct roles played by noise and measurement.  Furthermore, Wigner negativity is necessary for quantum computational advantage in continuous-variable settings~\cite{Mari_2012, Veitch_2012}, and the two descriptions used here assign it differently.  An observer who retains only the survival outcomes assigns the state to the ensemble $\rho^+(t)$, whose Wigner function is nonnegative for any passive initial state, and therefore carries no negativity resource.  By contrast, an observer who also retains the noise record assigns the states to an ensemble $P[\psi(t)]$, whose individual members contain Wigner negativity.  The negativity is thus not created by the choice of description but is accessible only to the observer holding the full record.  In this sense the first-passage protocol doubles as a heralded preparation scheme for non-Gaussian resource states, in which the only nonlinear element is the threshold measurement itself.

The protocol and model considered here are accessible in trapped-ion experiments~\cite{Ryan2025}, and these predictions can be tested directly.  This analysis should apply equally to bosonic modes in other platforms, such as single-mode photon fields where analogous measurements can be performed using so-called quantum scissors~\cite{scissors_1, scissors_2, scissors_3}, which have been used to truncate the Fock-space support of an optical field~\cite{Mattos_2022} and therefore produce quantum light from thermal light sources.

An interesting direction for future work is to determine whether measurement-induced nonclassical states can be exploited as a resource for sensing and metrology~\cite{mentesoglu2026sequentialmeasurementsresourcequantum}.  These schemes, when implemented on trapped-ion hardware, are naturally formulated as first-passage processes because photon recoil alters the motional state of interest, and therefore these protocols terminate when the first `bright' outcome is obtained.

\textbf{Author contribution statement} J.M.R. developed the theoretical formalism, performed the analytic calculations, and performed the numerical
simulations. J.M.R. and S.W.T. verified the analytical methods.  S.W.T. supervised the project. C.N. provided funding and supervised the project.  All authors contributed to scientific discussions that informed the interpretation of the results and contributed to the final manuscript.  ChatGPT and Claude were used for grammar checking, to assist in writing code that verified previously obtained results, and for the preparation of figures.

\textbf{Acknowledgements}
We are grateful to Jed Pixley for helpful discussions.  We also thank our group members Thomas J. Kessler and Mitchell G. Peaks for their support and many useful discussions.
This work was funded by NSF under QLCI: Center for Robust Quantum Simulation OMA-2120757.   J.M.R. was supported by the Goshaw Fellowship. 
\bibliography{bibliography}

\appendix
\clearpage
\widetext

\section{First-passage-time distribution}\label{app:QFPTD}
Using both the master equation and the quantum trajectories approach, we can solve for the first-passage-time distribution using numerical methods.  It is clear from Fig.~\ref{fig:escape_prob} that in the limit of $\theta\rightarrow 0$ the escape probabilities converge smoothly to a limit corresponding to continuous measurement.  The continuous measurement case is analytically tractable, and we now present some relevant results.

\begin{figure}
    \centering
    \includegraphics[width=0.5\linewidth]{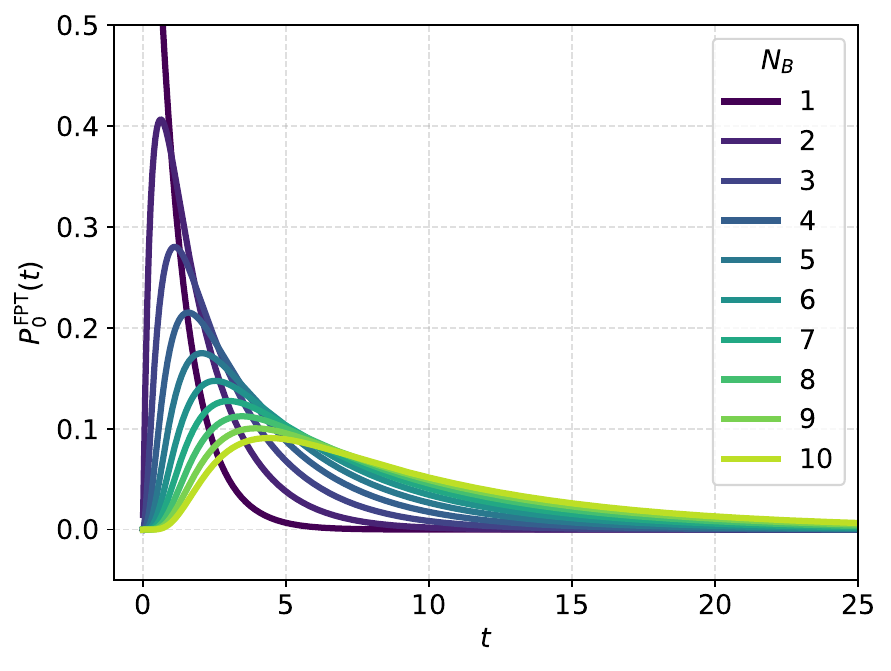}
    \caption{FPTD in the continuous measurement case ($\theta = 0$) for different values of $N_B$, and beginning in the ground state $\ket{0}$.}
    \label{fig:absorbing_fpt}
\end{figure}

As described in the main text, the dynamics on the surviving subspace obey~\eqref{eq:master_equation_Birth_death_main_text}, with an absorbing boundary condition $P_{N_B}(t) = 0$.  Note that the boundary at $n=0$ is already reflecting.  Use of this continuous-time approximation to the discrete-time renewal (\ref{eq:repeated_measurements}) can be justified by expanding it to first order in $\theta$ and then taking the $\theta\rightarrow 0$ limit.  The imposition of the absorbing boundary modifies the evolution compared to the case where there is no absorbing boundary (which corresponds to free evolution without measurement).  We now calculate both the quantum FPTD and escape probability in the continuous measurement case.  While the experimental measurement of these distributions~\cite{Ryan2025} only considered trajectories initialized in the ground state $\ket{0}$, here we consider arbitrary initial states.  We begin by denoting the escape probability as 
\begin{equation}\label{eq:survival}
E_{0}(t|N_0, N_B) = 1- \sum_{n=0}^{N_B-1} P_n(t)
\end{equation}
where, due to the absorbing boundary, $\sum_nP_n(t)\neq1$.  Furthermore, \eqref{eq:master_equation_Birth_death_main_text}~forms a closed set of ODEs.  Using Laplace transforms, we obtain a closed form for the escape probability in the Laplace domain $s$, without needing to know the individual $P_n(t)$.  We calculate $\lim_{t\to\infty}[P_n(t)]$ in Appendix~\ref{app:surviving_states}.  For a particle starting in $\ket{N_0}$ we obtain
\begin{equation}
    E_{0}(s|N_0, N_B) = \frac{\mathcal{L}_{N_0}(-s)}{s\mathcal{L}_{N_B}(-s)}.
\end{equation}By definition, the FPTD is the time derivative of the escape probability
\begin{equation}
    P^\text{FPT}_{0}(t|N_0, N_B) = \frac{dE_{0}(t)}{dt}.
\end{equation}
We then obtain
\begin{equation}\label{eq:fpt}
    P^\text{FPT}_{0}(s|N_0, N_B) =
\begin{cases}
  \frac{\mathcal{L}_{N_0}(-s)}{\mathcal{L}_{N_B}(-s)}, & N_0< N_B \\
  1, & N_0 \geq N_B
\end{cases}
\end{equation}
where ${\mathcal{L}_{i}(x)}$ is the $i$-th Laguerre polynomial.  From the poles of $P_0^\text{FPT}(s)$, we can transform (\ref{eq:fpt}) back to the time domain.  The poles of $P_0^\text{FPT}(s)$ correspond to the roots of ${\mathcal{L}_{N_B}(s)}$ which are all distinct, real and positive.  The long-time tail of the FPTD is of the form $P_0^\text{FPT}(t)\sim e^{-\lambda_{1}^{N_B}t} $ where $\lambda_{1}^{N_B}$ is the smallest root of $\mathcal{L}_{N_B}(x)$.  There is no known closed formula for $\lambda_{1}^{N_B}$, but it is known to be bounded by $1/N_B < \lambda_{1}^{N_B} <2/(N_B+1)$ for $N_B>1$ (for $N_B=1$, $\lambda_1^1 = 1$)~\cite{Gupta2007}.  The full FPTD in the time domain is then given by the Heaviside expansion theorem~\cite{ARFKEN2013963}
\begin{equation}\label{eq:app_fptd}
    P^\text{FPT}_{0}(t|N_0, N_B) = \sum_i\frac{\mathcal{L}_{N_0}(\lambda_i^{N_B})}{\mathcal{L}_{N_B-1}^{(1)}(\lambda_i^{N_B})}e^{-\lambda_i^{N_B}t}.
\end{equation}
Some examples of these continuous-time FPTDs in the time domain are shown in Figs.~\ref{fig:fptds_q_vs_c} and~\ref{fig:absorbing_fpt}.  For $N_B=1$, the FPTD is a pure exponential distribution $T\sim$~Exp~$(\lambda = 1)$, and does not have an initial ballistic part like those with $N_B>1$.  This effect is also present for finite $\theta$, where the FPTD for $N_B = 1$ is likewise a geometric distribution $T\sim$~Geo~$(p)$, where $p = \theta/(\theta + 1)$.  In both of these cases, this is because there is only one state in the surviving domain, and so every measurement whose outcome is \textit{survival} projects back into the ground state.  This is due to the quantization of energy levels, and does not occur `classically.'  The FPTD for an arbitrary initial state $\rho^i$ is given by $P_0^\text{FPT}(t) = \sum_n[\rho^i]_{nn}P_0^\text{FPT}(t| N_0 = n,N_B)$, which is a weighted average of the FPTDs over the initial state occupations.  The FPTD is independent of coherences in the initial density matrix.    

Since the Laplace transform of the FPTD is also its moment generating function, moments of the FPTD can be found.  For example, the mean first-passage time is given by:
\begin{equation}
\begin{split}
    M^\text{FPT}_1 &= -\frac{d}{ds}\bigg( P^\text{FPT}(s | N_0, N_B)\bigg)\bigg|_{s=0} \\
    &= N_B - N_0
\end{split}
\end{equation}
This result is consistent with the ensemble unit heating rate in our dimensionless units.  It is also consistent with the classical result, and therefore we see that the differences between quantum and classical theory do not appear in the first moment of the FPTD.

\section{Classical FPTD problem}\label{app:classical_problem}
In Appendix A we found a solution to the quantum first-passage-time problem in the $\theta = 0$ limit using a quantum master equation yielding~\eqref{eq:master_equation_Birth_death_main_text}.  However, the classical analogue of this first-passage process (i.e., (\ref{eq:classical_langevin})) is analytically challenging due to the multi-dimensional boundary condition, and analytical solutions remain elusive~\cite{Hanggi_classical_FPT, CRANDALL1970285}.  Inroads can be made using the stochastic averaging technique, wherein a second-order Markov process is reduced to a first-order process by changing the position-momentum variables $(x,p)$ into energy-phase variables $(E,\phi)$ using the Itô rule and Wong-Zakai correction terms.  This method assumes that the energy is constant over one period of oscillation, which is similar to the weak-coupling approximation in the derivation of the quantum master equation~(\ref{eq:master_eq}).  The resulting averaged stochastic differential equation is~\cite{VANVINCKENROYE2017328, VANVINCKENROYE2018178}:
\begin{equation}\label{eq:classical_ito}
    dE = dt + \sqrt{2E}dW.
\end{equation}
The moments of the classical FPTD $U_n(E_0, E_B) = \int t^nP_C^\text{FPT}(t)dt $ can be obtained recursively using the generalized Pontryagin equation~\cite{Gardiner}.  The first two moments have been calculated in the literature~\cite{VANVINCKENROYE2017328, VANVINCKENROYE2018178}.  Note that the classical and the quantum mean first-passage time are the same, so all the quantum–classical deviations live in the shape/higher moments of the FPTDs.  

The full first-passage-time distribution can be calculated from (\ref{eq:classical_ito}) by imposing a zero probability current condition at $E=0$ and an absorbing boundary at $E_B$, and using an eigenfunction expansion of the Fokker-Planck equation
\begin{equation}\label{eq:averaged_FPK}
    \frac{\partial P(E,t)}{\partial t} = -\frac{\partial P(E,t)}{\partial E} + \frac{1}{2} \frac{\partial^2}{\partial E^2}[2E P(E,t)].
\end{equation}
The absorbing boundary imposes $P(E_B) = 0$.  The expansion of (\ref{eq:averaged_FPK}) into eigenfunctions $Q_\lambda(E)$ with eigenvalues $\lambda$ yields a Sturm-Liouville problem
\begin{equation}\label{eq:sturm_liouville}
    E\frac{\mathrm{d}^2Q_\lambda}{\mathrm{d}E^2} + \frac{\mathrm{d}Q_\lambda}{\mathrm{d}E} + \lambda Q_\lambda = 0,
\end{equation}
with Bessel function solutions.  We then find that $\lambda_i = (j_0^{(i)})^2 / 4E_B$ where $j_0^{(i)}$ is the $i$-th root of the Bessel function of the first kind of order zero $J_0(x)$.  The eigenfunctions are then $Q_{\lambda_i}(E) = J_0(\sqrt{4\lambda_i E})$.  Thus, the full FPTD is
\begin{equation}
    P^\text{FPT}_{C}(t) = \sum_i A_{\lambda_i}(E_0, E_B) e^{- \lambda_i(E_B) t},
\end{equation}
where the coefficients $A_{\lambda_i}(E_0,E_B)$ are determined by the projection of the initial condition onto the eigenfunctions $Q_i(E)$ under the appropriate Sturm-Liouville normalization.  Unlike in the quantum case where there are a finite number of discrete eigenvalues, in the classical case there are an infinite number of such discrete eigenvalues.  Furthermore, we note that the Mehler-Heine formula $\lim_{n\rightarrow\infty} \mathcal{L}_{n}(z^2 / 4n) = J_0(z)$ gives an obvious connection between the quantum solution in terms of Laguerre polynomials, and the classical solution in terms of Bessel functions.  

In an analogous fashion, we find the quasi-stationary phase-space density of the classical surviving states
\begin{equation}\label{eq:classical_phase_space_density}
    P(r = \sqrt{2E}) =
\begin{cases}
  \kappa J_0(j_0 r / \sqrt{2E_B}), & r<r_B \\
  0, &  r\geq r_B,
\end{cases}
\end{equation}
where $\kappa = j_0 / (4\pi E_B J_1(j0))$ is a normalization factor and where $j_0\approx 2.40483$.

\section{Surviving states}\label{app:surviving_states}
\begin{figure}
    \centering
    \includegraphics[width=0.5\linewidth]{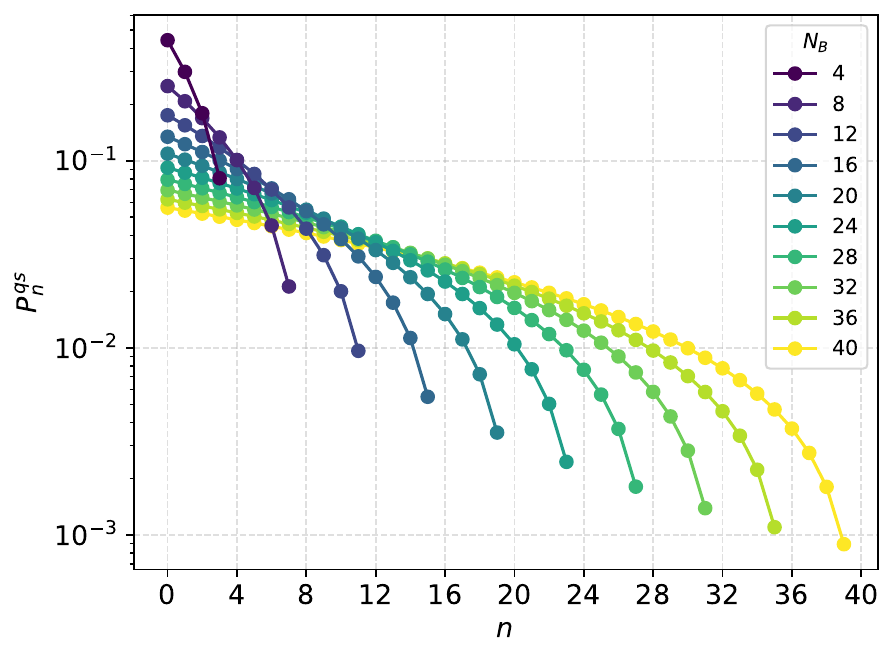}
    \caption{Quasi-stationary state oscillator occupation probability distribution $P_n^\text{qs}(N_B)$ for $\theta = 0$, as calculated in (\ref{eq:steady_state}) and (\ref{eq:steady_state_appendix}).}
    \label{fig:quantum_steady_state_distribution}
\end{figure}

At long times the occupation probabilities in~\eqref{eq:master_equation_Birth_death_main_text} approach a quasi-stationary state: $P_n(t)\propto e^{-\lambda_{1}^{N_B}t}P_n^{qs}$, where $P_n^{qs}$ are the normalized quasi-stationary state occupation probabilities and where $\lambda_{1}^{N_B}$ is the smallest Laguerre root.  Using this ansatz we find $P_n^{qs}$ by eliminating the time dependence from~\eqref{eq:master_equation_Birth_death_main_text}.  Accordingly, the quasi-stationary state, which is the eigenmode of~\eqref{eq:master_equation_Birth_death_main_text} subject to the absorbing boundary associated with the smallest eigenvalue $-\lambda_{1}^{N_B}$, satisfies
$$\mathbf{A}\vec P^{qs} = \mathbf{0}$$
where
\begin{equation}\label{eq:A_tridiag}
\mathbf{A} =
\begingroup\setlength\arraycolsep{3pt}
\begin{bmatrix}
\lambda-1 & 1 & 0      & \cdots & 0 \\
1 & \lambda-3 & 2  & \ddots & \vdots \\
0     & 2 & \lambda-5  & \ddots & 0 \\
\vdots& \ddots& \ddots & \ddots & N_B-1 \\
0     & \cdots& 0      & N_B-1  & \lambda-2N_B+1
\end{bmatrix}
\endgroup
\end{equation}
where $\lambda= \lambda_{1}^{N_B}$.  The quasi-stationary state is the kernel of $\mathbf{A}$, i.e., the eigenvector of the generator $\mathbf{A} - \lambda\mathbf{1}$
with eigenvalue $\lambda = -\lambda_1^{N_B}$. Since $\mathbf{A}$ is a Jacobi matrix, its eigenvalues are simple, so this eigenvector, and hence the quasi-stationary state, is unique.  We then find the kernel of $\mathbf{A}$ by solving a three-term recurrence and obtain
\begin{equation}\label{eq:steady_state_appendix}
    P_n^{qs} = \frac{1}{C}\mathcal{L}_n(\lambda)
\end{equation}
where we include a normalization factor $C = \sum_{n=0}^{N_B-1}\mathcal{L}_n(\lambda) = \mathcal{L}_{N_B}^{(1)}(\lambda_1^{N_B})$.  By normalizing $P_n^{qs}$ we are calculating the occupation probability of the \textit{surviving} states only.  We show these occupation probabilities for different $N_B$ in Fig.~\ref{fig:quantum_steady_state_distribution}.  Notice that in contrast with a thermal state, which satisfies the same dynamical equation except of course without the absorbing boundary and has an exponential decay over $n$, the absorbing boundary now gives a superexponential decay over $n$.  

We note that the conditioned population dynamics~\eqref{eq:master_equation_Birth_death_main_text} preserve passivity.  An oscillator state $\rho_p$ is called passive if $\rho_p$ is diagonal in the Fock basis, and if $p_n \geq p_{n+1}$ $\forall n$.  Thermal states are examples of passive states.  These states are called passive as no work can be extracted from them by any unitary acting only on the oscillator. Suppose the populations are non-increasing, $P_0\geq P_1 \geq \dots \geq P_{N_B - 1}$, with $P_{n} = P_{n+1}$ for some $n$. Then~\eqref{eq:master_equation_Birth_death_main_text} gives $\partial_t(P_n - P_{n+1}) = n(P_{n-1} - P_n) + (n+2)(P_{n+1} - P_{n+2}) \geq 0$, so the ordering cannot be violated: the set of passive states remains passive under time evolution. Since passive states have nonnegative Wigner functions~\cite{Herstraeten_2021}, the ensemble-averaged surviving state derived from any passive initial state, including the ground state and thermal states, has a nonnegative Wigner function at all times, and its Wigner negativity vanishes identically.

\section{Non-Gaussianity of the surviving states}\label{app:nonGaussianity}
\begin{figure}[h!]
    \centering
    \includegraphics[width=0.5\linewidth]{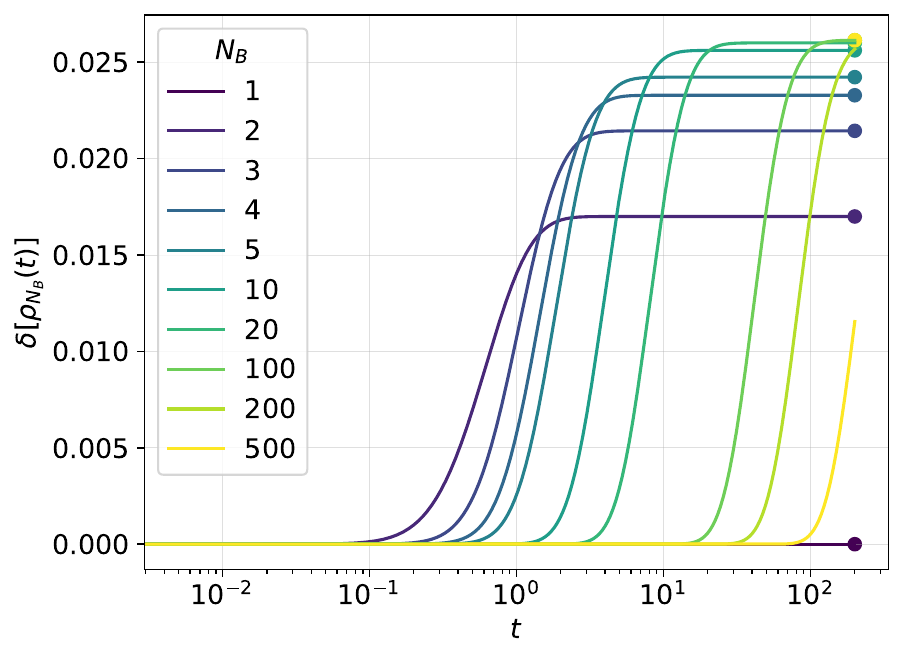}
    \caption{Non-Gaussianity $\delta[\rho_{N_B}(t)]$ of the surviving motional states for $\theta = 0$ as a function of time $t$ for different $N_B$.  The dots indicate the non-Gaussianity of the analytical steady state of (\ref{eq:steady_state}), which are formally for $t\rightarrow\infty$, whereas the lines are generated by numerically solving~\eqref{eq:master_equation_Birth_death_main_text} and then evaluating the non-Gaussianity.  }
    \label{fig:non-gaussianity}
\end{figure}
We quantify the degree of non-Gaussianity of the surviving states using a non-Gaussianity measure $\delta[\rho]$~\cite{Banazsek2007}.  It is based on the square Hilbert-Schmidt distance between the state $\rho$ and a reference Gaussian state $\rho_G$ which depends on $\rho$ and has the same covariance matrix.  In our case, $\rho_G$ is a thermal state with the same mean occupation as $\rho$.  The non-Gaussianity is defined as
\begin{equation}
    \delta[\rho] = \frac{D_\text{HS}[\rho, \rho_G]}{\mu[\rho]}
\end{equation}
where $D_\text{HS}[\rho, \rho_G] = \frac{1}{2}\Tr [(\rho-\rho_G)^2]$ is the square Hilbert-Schmidt distance and $\mu(\rho) = \Tr [{\rho}^2]$ is the purity.  We calculate the non-Gaussianity of the surviving states in the continuous measurement case as a function of time $\delta[\rho_{N_B}(t)]$ in Fig.~\ref{fig:non-gaussianity}.  Interestingly, for large $N_B$, $\delta[\rho_{N_B}(t\rightarrow\infty)]$ approaches a limiting value indicating that the shape differences causing the non-Gaussianity saturates.  This corresponds to the classical non-Gaussianity.  Furthermore, the non-Gaussianity is initially zero, and remains there until a time $t\sim N_B$ where it increases rapidly until it reaches a quasi-stationary state.  This shows that initially, the measurements do not modify the thermal evolution substantially until such a time as there is substantial occupation near the boundary whereupon the measurements substantially modify the surviving states.

In order to illustrate the non-Gaussianity, we numerically evaluate the steady-state Wigner function in the continuous measurement limit and show the results in Fig.~\ref{fig:wigner}.  The steady-state Wigner functions are visibly non-Gaussian~\cite{Walschaers2021} and, for small $N_B$, display oscillatory features absent in the classical phase-space distribution.  The ensemble Wigner function of the quasi-stationary state distribution (\ref{eq:steady_state}) shows oscillations for small $N_B$, which progressively disappear for larger $N_B$.  The phase-space distribution of the classical motional quasi-stationary state does not show these oscillations.  It is no surprise that when $N_B$ is small, and thus the number of surviving phonons is small, quantum effects are important and a classical description is inadequate.

\section{Monte Carlo simulation}\label{app:MonteCarlo}
\begin{figure}[h!]
    \centering
    \includegraphics[width=0.5\linewidth]{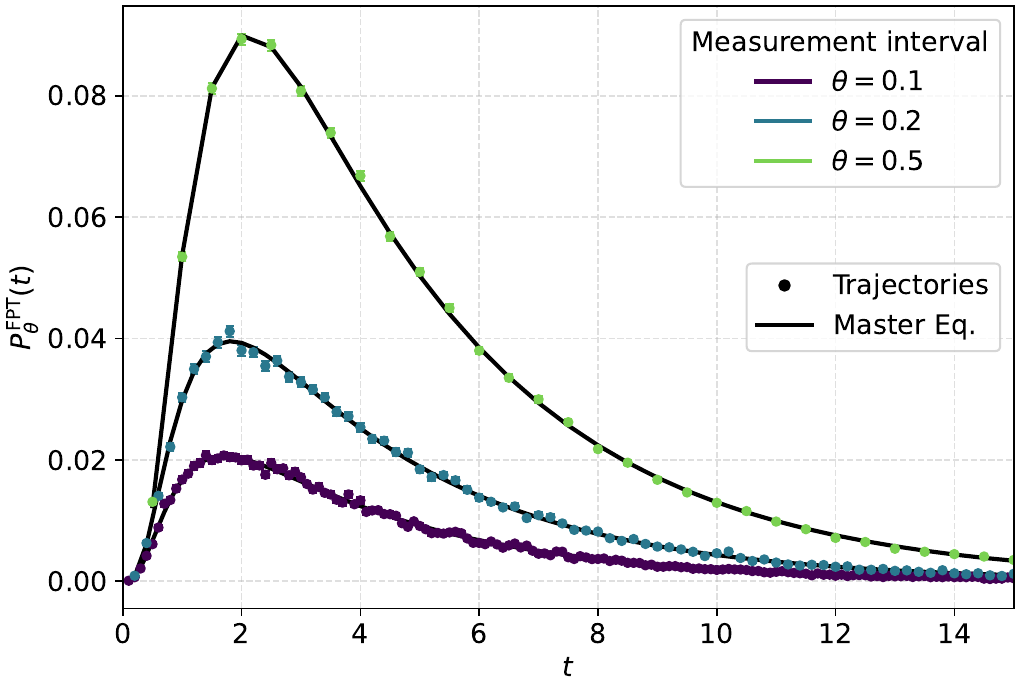}
    \caption{Comparison between the FPTDs ($N_0 = 0, N_B=4$) obtained using master equation methods (black solid lines) and the corresponding FPTDs obtained using the Monte Carlo trajectories simulation (colored dots with $1\sigma$ error bars reflecting the finite sample size).  }
    \label{fig:trajectories_comparison_to_master}
\end{figure}

We simulate the process in (\ref{eq:displaced}-\ref{eq:traje_2})  using a Monte Carlo method.  We generate $5\times10^4$ individual trajectories.  We compare the distribution of stopping times obtained from the trajectories simulation to those obtained from the deterministic integration of the master equation stroboscopic evolution (\ref{eq:master_eq}-\ref{eq:def_FPTD}) in Fig.~\ref{fig:trajectories_comparison_to_master} where we observe excellent agreement between the two.  We observe similar agreement over a broad range of starting and barrier energies.

\end{document}